\documentclass[trackchanges,twocolumn]{aastex631}

\usepackage{graphicx}
\usepackage{dcolumn}
\usepackage{bm}
\usepackage{color}
\usepackage{amsmath}

\begin{document}

\title{Three dimensional, spherically polarized magnetic fields}%

\author{Anna Tenerani}
 \email{Anna.Tenerani@austin.utexas.edu}
\affiliation{%
 Department of Physics, The University of Texas at Austin, TX 78712, USA
}%

\author{Marco Velli}
\affiliation{
 Department of Earth, Planetary, and Space Sciences, University of California, Los Angeles, CA 90095, USA
}%

\begin{abstract}
Turbulence in the solar wind is characterized by Alfv\'enic fluctuations that exhibit spherical polarization, a geometric condition resulting in the nearly constant magnitude of the magnetic field. This property persists even during the largest field fluctuations, sometimes leading to local polarity reversals known as switchbacks. 
A longstanding question is whether three-dimensional smooth magnetic fields can simultaneously satisfy the constant-$|{\bf B}|$ constraint, and how such fields can be constructed analytically or numerically. Here we propose a new numerical method that allows to construct a magnetic field that is exactly spherically polarized, reproducing key features of solar wind fluctuations. Using this framework, we find evidence that discontinuities are  unavoidable for generic three-dimensional configurations. Fundamentally, this implies that field rotations cannot maintain exactly constant $|{\bf B}|$ in an arbitrarily large spatial domain. Rather, field rotations with constant magnitude can exist in limited regions of space.  We argue that these finite spatial domains are separated by discontinuities where a local departure from constant $|{\bf B}|$ is expected. These results provide insights into the structure of solar wind turbulence and more generally into the nature of nonlinear magnetic fluctuations in plasmas.
\end{abstract}



\section{Introduction}
The solar wind emerging from coronal holes displays signatures of Alfvénic turbulence, with magnetic and velocity fluctuations that remain remarkably coherent over large distances and throughout the inertial range of the turbulent cascade. These fluctuations are characterized by both a high degree of Alfvénic correlation (cross helicity $\lesssim 1$) \citep{belcher1971large,bruno2005observations,chen2020evolution,shi2021alfvenic} and a well‑defined phase correlation among the fluctuating field components resulting in spherical polarization \citep{barnes1974large,tsurutani2018review,matteini2015ion}. This geometrical property manifests as a nearly constant magnetic field magnitude $|{\bf B}|$, even during large‑amplitude magnetic fluctuations leading to polarity reversals (switchbacks) regularly observed by Parker Solar Probe (PSP) \citep{kasper2019alfvenic,bale2019highly}. 
%
Explaining how large field rotations are achieved dynamically while preserving an approximately constant $|{\bf B}|$ has become a central problem in solar wind turbulence. One longstanding question is whether smooth three‑dimensional (3D) magnetic field configurations can satisfy simultaneously the solenoidal constraint or if discontinuities are a fundamental element of such field rotations. Smooth magnetic field fluctuations with exact constant magnitude can be defined analytically in the case of plane waves in parallel \citep{malara1996parametric} or oblique propagation (thus 1D) \citep{marriott2024parametric}, in 2D \citep{primavera2019parametric,tenerani2020magnetic}, or in 3D but with some symmetry such as axisymmetry \citep{shi2024analytic}. This problem, however, is nontrivial mathematically and numerically in a truly 3D field, and carries important implications for understanding the geometry and topology of switchbacks and, more generally, the structure of solar wind turbulence. Although different numerical methods have been proposed to build magnetic field configurations with constant $|{\bf B}|$ \citep{roberts2012construction,valentini2019building,squire2022construction,huang2025switchbacks}, a clear answer to this question has remained elusive. 
Existing approaches are based on phase optimization or other relaxation method in Fourier space to minimize fluctuations of $|{\bf B}|$ while enforcing solenoidality \citep{roberts2012construction,huang2025switchbacks}, numerical integration of the vector potential under the constant-$|{\bf B}|$ constraint \citep{valentini2019building}, or amplitude growth through an induction equation \citep{squire2022construction}. These methods produce magnetic fields with approximately constant $|{\bf B}|$ that typically exhibit sharp field variations. However, numerical resolution limits their unambiguous identification, making a rigorous interpretation of discontinuity formation problematic \citep{squire2022construction}. Additionally, these numerical methods have been shown to work in reduced dimensionality \citep{roberts2012construction} or to struggle with generating large mean-field-aligned rotations \citep{valentini2019building}. In some cases, they inevitably generate sharp discontinuities at switchback edges \citep{huang2025switchbacks} or require a small amplitude fluctuating constant-$|{\bf B}|$ field as  initial condition~\citep{squire2022construction}.   

In this work, we propose a new method to find (solenoidal) magnetic fields satisfying the constant-$|{\bf B}|$ condition, where this constraint is enforced exactly by representing the field through phase angles. Our approach is different from that adopted by \citet{goldstein1974}, where phase angles are used to describe the fluctuation, rather than the total magnetic field. Furthermore, that formulation leads to an ambiguous definition of the unperturbed magnetic field, which can suggest the existence of plane-wave solutions that are localized along the mean-field direction while maintaining constant $|{\bf B}|$. However, \citet{barnes1976nonexistence} later demonstrated the nonexistence of such solutions. Because our method can also be treated semi‑analytically, it  enables a more general investigation into questions of existence, regularity, and smoothness for constant-$|{\bf B}|$  magnetic fields in three dimensions. 

\section{Method}

We represent the magnetic field as a rotation on a sphere of constant radius $B$,
\begin{equation}
    {\bf B} = B(\cos \theta\,\hat x+ \sin \theta \sin \phi\,\hat y+\sin \theta \cos \phi\, \hat z),
\end{equation}
%
where $\theta$ and $\phi$ are functions of the spatial coordinate ${\bf x}=(x, y, z)$, and we fix $B=1$. The solenoidal condition  then yields 
%
\begin{equation}
\begin{split}
     \partial_z \phi = -\frac{1}{\sin\phi}\partial_x \theta + &  {\frac{\cos\theta}{\sin\theta}}\partial_y \theta \\
     &+ {\frac{\cos\theta}{\sin\theta}}{\frac{\cos\phi}{\sin\phi}}\partial_z \theta + \frac{\cos\phi}{\sin\phi}\partial_y \phi.
\end{split}
\label{pd_phi_z}
\end{equation}
%
For given smooth $\theta({\bf x})$ and initial condition $ \phi_0(x,y)=\phi(x,y,0)$, eq.~(\ref{pd_phi_z}) is a well posed PDE that can be integrated numerically for the phase $\phi({\bf x})$, provided $\sin\theta\neq0$ and $\sin\phi\neq 0$. Thus, it is possible to reproduce rotations covering half-sphere, with $0<\theta<\pi$ and $0<\phi<\pi$. A smaller or larger angular coverage means smaller or larger amplitude of field fluctuations. Here we use an explicit 3rd order Runge-Kutta scheme with periodic boundary conditions in $x$ and $y$ to integrate eq.~(\ref{pd_phi_z}) forward starting from a reference plane that  we fix at $z=0$. 

To  first gain insight into the structure of eq.~(\ref{pd_phi_z}), it is convenient to formulate it as a system of ODEs with the method of characteristics,
\begin{equation}
    \frac{dy}{dz} = - u(\phi), \qquad \frac{d\phi}{dz} = F(x,y(z),z,\phi),
    \label{char}
\end{equation}
where we have defined
\begin{equation}
      u(\phi)=\frac{\cos\phi}{\sin\phi}, 
    \label{eq:u_def}
\end{equation}
\begin{equation}
\begin{split}
      F(x,y,z,\phi) =& -\frac{1}{\sin\phi}\partial_x \theta + {\frac{\cos\theta}{\sin\theta}}\partial_y \theta \\
      &+ {\frac{\cos\theta}{\sin\theta}}{\frac{\cos\phi}{\sin\phi}}\partial_z \theta.
\end{split}
    \label{eq:F_def}
\end{equation}
%
For smooth and continuously differentiable $F$, a continuous solution to eqs.~\eqref{char} is known to exist in the neighborhood of the initial condition. However, in many cases the characteristics $y(z)$ intersect, leading to genuinely discontinuous solutions at a finite distance $z$. Thus, the existence of a smooth solution in an arbitrarily large domain cannot be guaranteed. When this happens, the conditions of constant $|{\bf B}|$ and $\boldsymbol\nabla \cdot {\bf B}=0$ cannot be satisfied simultaneously.  This suggests  that a local relaxation of the constant-$|{\bf B}|$ constraint is generally required.

Owing to the nonlinear forcing term $F$, eqs.~(\ref{char}) must, in general, be solved numerically. However, there are particular choices of $\theta({\bf x})$ for which analytic or semi-analytic solutions can be found, highlighting some general conditions under which discontinuities in the solution form.  To illustrate this point, we discuss below two cases (case 1 and case 2) that satisfy the minimal requirements for a 3D field where $\theta=\theta(x)$ and $\theta=\theta(y)$, respectively. For these cases, characteristic intersections can be analyzed for generic initial conditions.  
We then extend our method to construct a  broadband spectrum with a 3D phase $\theta(x,y,z)$ (case 3) similar to solar wind observations. For this case, no comparable general result is available, and we rely on explicit numerical evidence that suggests that intersections occur for generic configurations.

\subsection{Case 1} We consider the special case $\theta=\theta(x)$, for which eqs.~(\ref{char}) are analytically integrable along characteristics. Integrating the second equation from $z=0$ with initial condition $\phi_0(x,y)$ and substituting into the first yields the characteristic curves $y(z)$: 
\begin{equation}
\label{eq:char-explicit}
y(z;y_0\,|\,x)=y_0+\frac{1}{a}\left[\sqrt{1-(az+D)^2}-\sqrt{1-D^2}\right].
\end{equation}

In the equation above, we label each characteristic by its footpoint $y(0)=y_0$ and we define
\begin{equation}
a=\theta'(x), \qquad D=\cos\phi_0(x,y_0).
\end{equation}
%
The characteristics  form a family of semicircles in the $(y,z)$ plane whose center and radii of curvature depend on the footpoint $y_0$. As a consequence, these curves eventually intersect, implying the formation of genuine discontinuities in $y$ at finite $z$ for fixed $x$. Alternatively, the solution becomes singular when $az+D=1$ before intersection occurs. In either case, smooth three-dimensional solutions with exactly constant $|{\bf B}|$ do exist over a finite interval in $z$, but discontinuities (or singularities) eventually develop. The extent of the interval where a smooth solution exists increases as  $\theta^\prime(x)$ decreases, that is, as the amplitude of the field rotation becomes smaller or if it varies slowly. 
\begin{figure}
    \centering
    \includegraphics[width=0.6\linewidth]{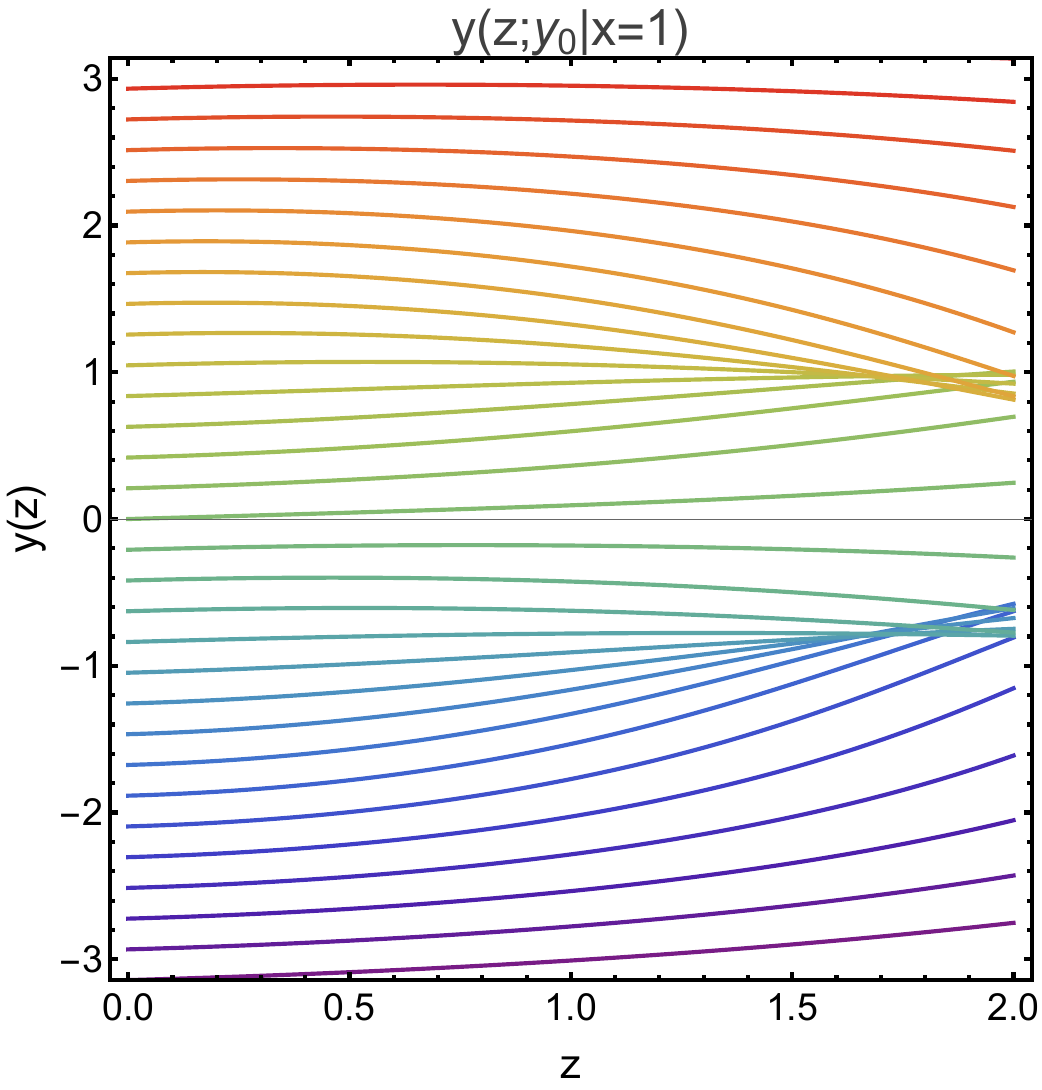}
    \caption{Case 2: A family of characteristic curves  at $x=1$, showing their intersection at $z\simeq 1.7$. Colors indicate a different footpoint $y_0$.}
    \label{Switchback_across_char}
\end{figure}
\subsection{Case 2} When the phase $\theta=\theta(y)$, eqs.~(\ref{char}) can be combined, yielding
\begin{equation}
 \sin\phi=-\left(\frac{\sin\phi_0(y_0|x)}{\sin\theta(y_0)} \right)\sin\theta(y).
\end{equation}
The characteristics are therefore determined by solving
\begin{equation}
    \frac{dy}{dz}=\mp\frac{\sqrt{1-(\sin\phi)^2}
    }{\sin\phi}.
    \label{char_across}
\end{equation}
The form of the right-hand-side of eq.~\eqref{char_across} shows that the characteristic velocity depends on both $x$ and the footpoint $y_0$ through $\phi$. Consequently, neighboring characteristics generally propagate at different speeds, providing a mechanism for characteristics convergence and eventual intersection.  For $z\ll1$, the characteristics are approximated to lowest order by $y(z)\simeq y_0-z/\tan(\phi_0(y_0,x))$. Because intersections occur when $dy/dy_0=0$, an estimate for the crossing distance is $z_c\simeq-\sin^2\phi_0/\phi_0^\prime$.  If $\phi_0^\prime=0$, the dominant linear term vanishes and a higher order expansion is necessary.  Expanding eq.~\eqref{char_across} to the next non-vanishing order shows that neighboring characteristics start to converge when $\theta^{\prime\prime}\cot\theta-(\theta^\prime)^2\csc^2\theta>0$ with an estimated $z_c\simeq\pm\sqrt{(2 (\sin\phi_0)^2)/(\theta^{\prime\prime}\cot\theta-(\theta^\prime)^2\csc^2\theta})$. Since  $\theta$ and $\phi_0$ are oscillatory within the hemisphere, they necessarily contain regions where neighboring characteristics are compressive. The estimates above therefore predict  finite crossing distance $z_c$, indicating that intersection of characteristics is generically expected. 
As an example, we construct a field rotation reproducing a switchback by choosing
\begin{equation}
    \theta(y) = \frac{1}{2}+\tanh(y+1)-\tanh(y-1),
    \label{theta_2}
\end{equation}
\begin{equation}
    \phi_0(x) = \frac{\pi}{2} +0.1\sin\left(x \frac{2\pi}{L_x}\right),
    \label{phi_2}
\end{equation}
in the range  $-L_x/2\leq x\leq L_x/2$, $-L_y/2\leq y\leq L_y/2$ with $L_x=L_y=2 \pi$.  
Figure~\ref{Switchback_across_char} shows a family of characteristics for this system at $x=1$ that converge and intersect at $y\simeq \pm 1$ and $z\simeq1.7$ where the solution breaks down. As in case~1, the extent of the domain of existence becomes larger for smaller amplitude fluctuations. 

\begin{figure}
    \centering
            \includegraphics[width=0.8\linewidth]{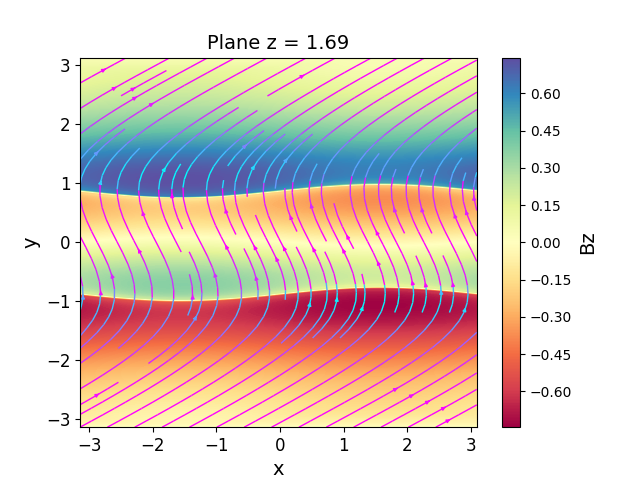}
                \includegraphics[width=0.8\linewidth]{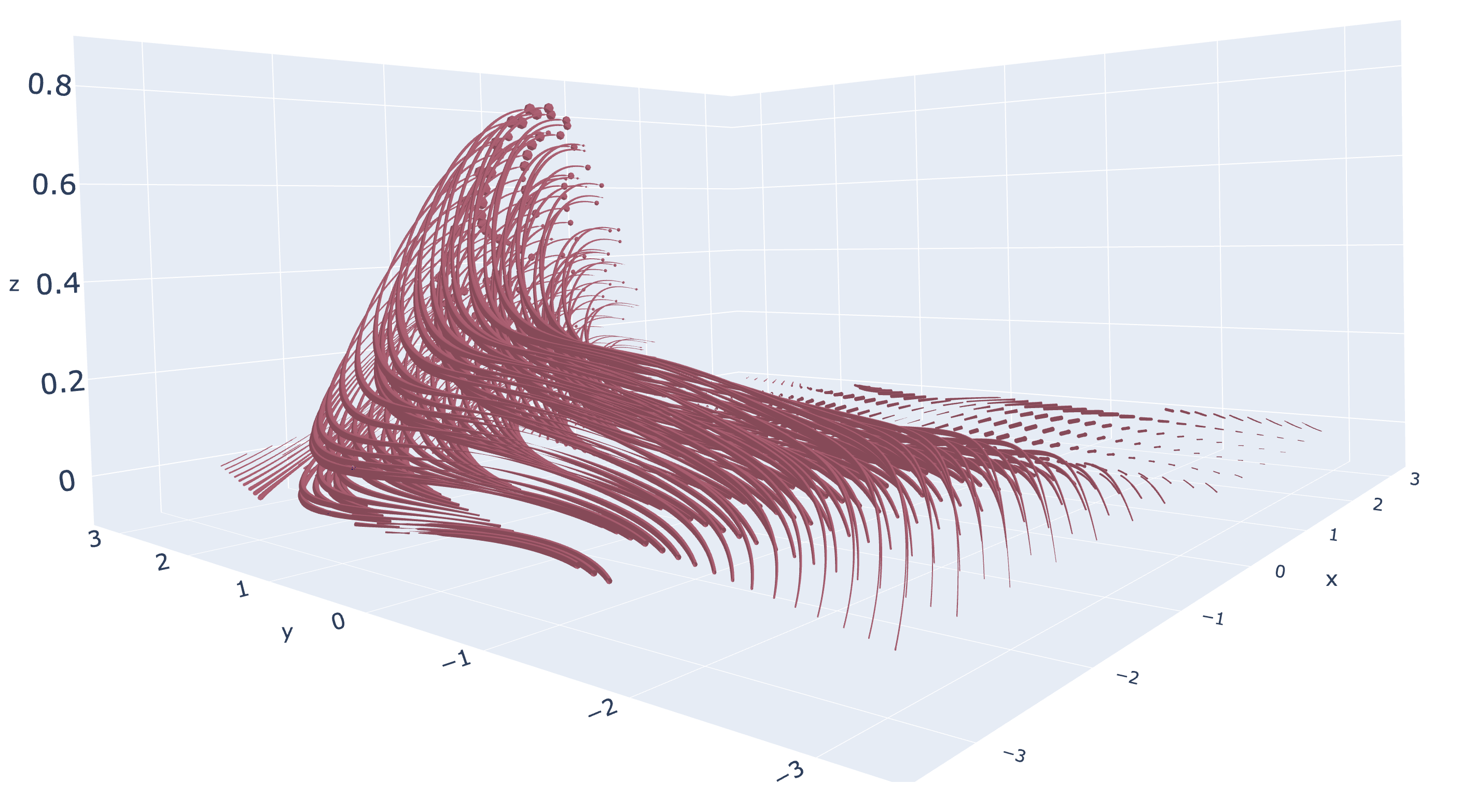}
    \caption{Case 2. Top panel: magnetic field lines projected in the $(x,y)$ plane at $z=1.69$. $B_z$ is color coded as indicated in the side bar. Bottom panel: magnetic field lines in 3D space.}
    \label{Switchback_across}
\end{figure}

The magnetic field shown in Fig.~\ref{Switchback_across}  was reconstructed after integration of eq.~\eqref{pd_phi_z}. Achieving good numerical accuracy in this integration (with $\nabla\cdot\mathbf B \ll 1$ numerically) becomes increasingly challenging as the phase $\phi$, and hence $\mathbf B$, steepens. However, the method of characteristics informs on the domain in which regular solutions exist, ensuring that the numerical solution is physically correct despite increasing numerical errors on the divergence. The integration was therefore performed on a domain $L_x\times L_y\times L_z = 2\pi\times2\pi\times1.69$ using $128\times512\times256$ mesh points with a pseudospectral filter in the $y$ direction to prevent numerical instabilities as the steepening develops.
%
%
The resulting unperturbed field is primarily along $x$, ${\bf B} = (0.86,0.5,,0)$, with rms fluctuations $\delta B_x = 0.47$, $\delta B_y = 0.18$, and $\delta B_z = 0.23$. A developing discontinuity at $y\simeq\pm1$ is apparent in the contour plot of $B_z$ at the plane $z=1.69$ (Fig.~\ref{Switchback_across}, top panel), and is accompanied by a sharp deflection of $B_y$ at the same location. The resulting field lines are shown in the full 3D domain in Fig.~\ref{Switchback_across}, bottom panel. 
This explicit example illustrates that switchbacks do not need to exhibit sharp boundaries everywhere a field reversal exists, although discontinuities eventually develop. 
Additionally, a consequence of our construction method is that while the switchback can be localized in the $(x,y)$ plane, it necessarily extends in the third ($z$) direction. Thus, consistent with the analytic results of \citet{shi2024analytic}, our example indicates that a switchback can be fully bounded in three dimensions only if the constant-$|\bf{B}|$ condition is relaxed.   

\subsection{Case 3: solar wind-like deflections}
We build a broadband spectrum starting from a random superposition of scale-independent rotations:
\begin{equation}
\begin{split}
      \tilde\theta({\bf x}) = \sum_{k_x,k_y,k_z}&\cos\left(k_x x +\varphi_{k_x}\right) \\
    &\cos\left(k_y y +\varphi_{k_y}\right)
      \cos\left(k_z z +\varphi_{k_z}\right),
          \label{theta_broadband}
\end{split}
\end{equation}
\begin{equation}
    \phi_0(x,y) = \sum_{k_x,k_y}\cos(k_x x +\varphi_{k_x})\cos(k_y y +\varphi_{k_y}),
    \label{phi_broadband}
\end{equation}
where we have used the first 5 modes in each direction and $\varphi_i$ are random phases. We then renormalize the extrema of $\theta$ and $\phi_0$ to prescribe the desired angular coverage of the field, which controls the amplitude of magnetic fluctuations. The dependence of $\theta$ on $z$ now modifies the behavior of characteristics discussed in previous paragraphs, and a general conclusion on existence of smooth solutions cannot be drawn. Nevertheless, due to the nonlinear and oscillatory behavior of $F$, it is expected that discontinuities form at some $z$ for generic phase functions. As in the previous cases, we empirically find that the admissible range of domain sizes $L_z$ expands as the fluctuation amplitude decreases, or if wavelengths increase. 

The phases defined in eq.~\eqref{theta_broadband}-\eqref{phi_broadband} produce angle distributions that are approximately Gaussian and centered near the midpoint of the imposed range. Consequently, large deflections with $\theta>\pi/2$ lack the intermittent character typical of solar‑wind switchbacks. 
To generate sparse large‑amplitude deflections in $B_x$, we apply an exponential mapping to produce a skewed distribution:
\begin{equation}
    \theta = \theta_0+\beta\left( e^{\alpha \tilde\theta} -min(e^{\alpha \tilde\theta})\right),
\end{equation}
where $\theta_0$ sets the smallest value of $\theta$, and $\beta$ and $\alpha$ control the skewness and variance. In the case-study shown here we chose $\theta_0=0.1$, $\alpha=5$ and $\beta=0.1$. 
We integrated eq.~\eqref{pd_phi_z} in a domain $L_x=L_y=2\pi$ and $L_z=0.3$ 
using $128\times512\times256$ mesh points. 
\begin{figure}
    \centering
            \includegraphics[width=0.8\linewidth]{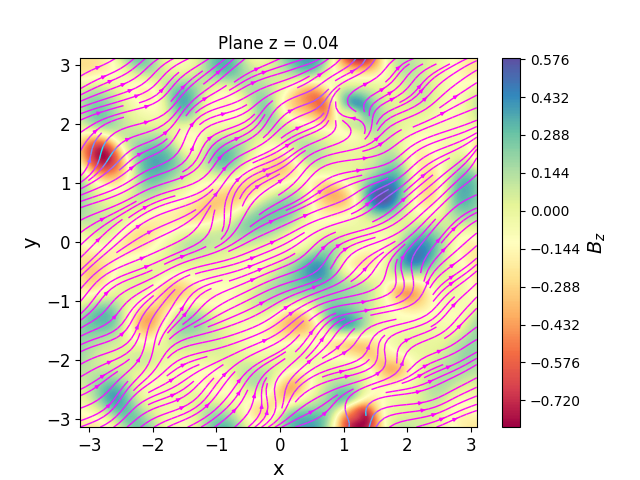}
            \includegraphics[width=0.8\linewidth]{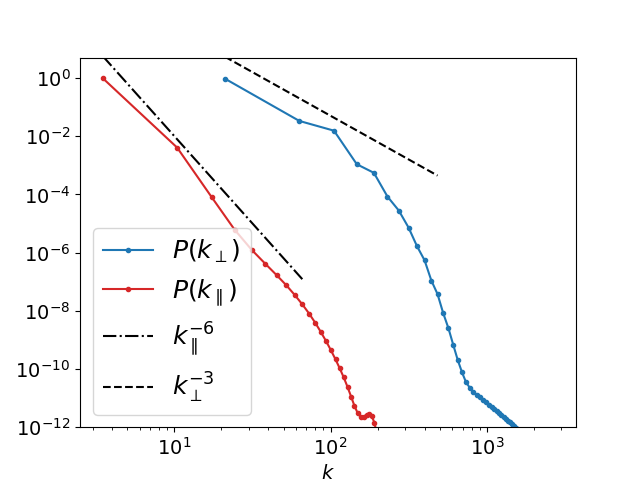}
    \caption{Case 3. Top panel: magnetic field lines projected in the $(x,y)$ plane at $z=0.0355$ with $B_z$ in color code as indicated in the side bar. Bottom panel: reduced power spectrum as a function of $k_\perp$ (blue) and $|k_\parallel|$ (red); dashed and dot-dashed lines indicate power laws of $k_\perp^{-3}$ and $k_\parallel^{-6}$ for reference.}
    \label{broadband}
\end{figure}
\begin{figure}
    \centering
    \includegraphics[width=0.8\linewidth]{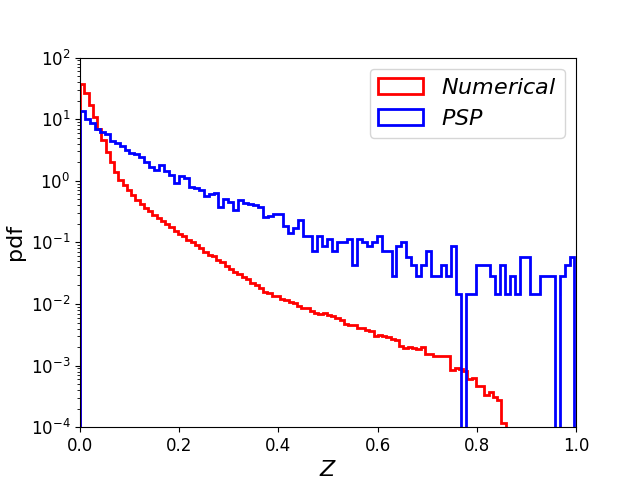}
    \includegraphics[width=0.8\linewidth]{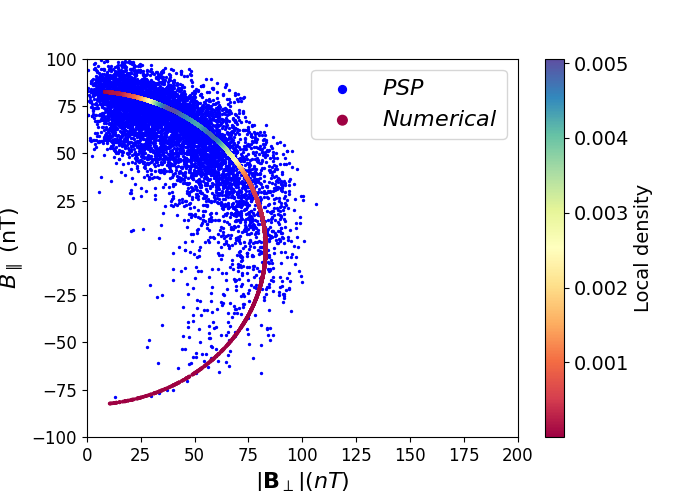}
        \includegraphics[width=\linewidth]{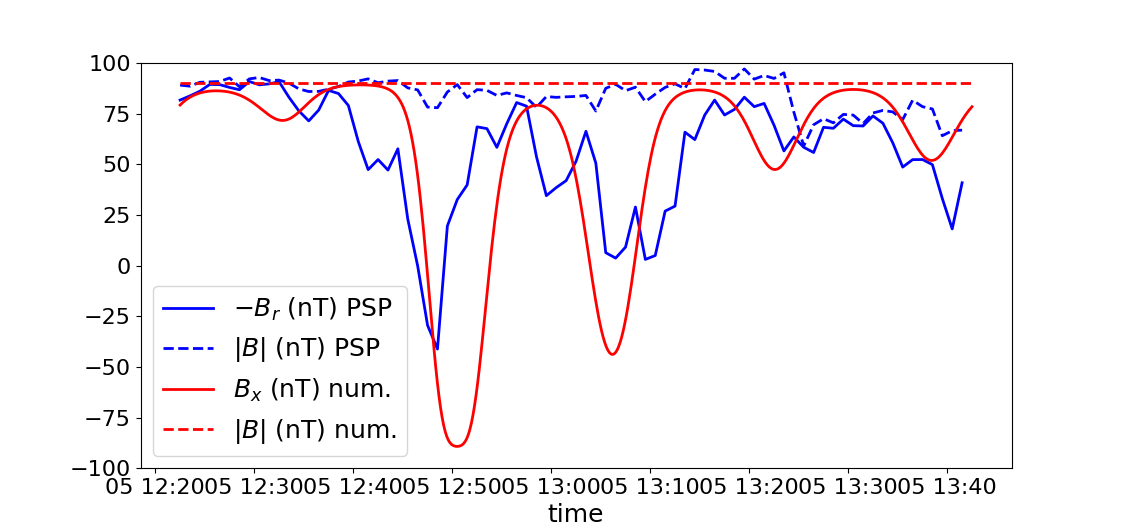}
    \caption{Case 3. Comparison between numerical solution (red) and PSP data (blue). Top panel: distribution of the deflection parameter $\mathcal{Z}$. Middle panel: scatter plot of magnetic field components. Bottom panel: comparison between a 1d cut of the numerical solution and a time sequence  of PSP data.}
    \label{comparison}
\end{figure}
The reconstructed magnetic field has a mean $\langle{\bf B}\rangle=(0.75,0.57,-0.006)$ and rms amplitudes $\delta B_x=0.2$, $\delta B_y=0.18$ and $\delta B_z=0.178$ with embedded switchbacks. The largest deflection is reached at ${\bf x}=(1.13,2.12,0.0355)$ where $B_x=-0.99$. Fig.~\ref{broadband}, top panel, shows the magnetic field at $z=0.0355$, where a switchback can be seen near the top right corner of the plot. The bottom panel shows the 1D reduced spectrum of the resulting field as a function of the perpendicular and parallel wave number, $k_\perp$ and $|k_\parallel|$, defined with respect to the mean field. The resulting spectrum is steeper than observed, scaling approximately as $k_\perp^{-3}$ and $|k_\parallel|^{-6}$, although it displays an anisotropic distribution with energy mostly in $k_\perp$ as in the solar wind \citep{dasso2005anisotropy,horbury2012anisotropy,sioulas23}.  We note that we tested different initial conditions with weighted Fourier coefficients and by increasing the number of modes, and we found that the resulting magnetic spectra remain largely unaffected. 

For the sake of illustration, Fig.~\ref{comparison} presents a comparison between the numerical solution and PSP data at 1 minute resolution during the first perihelion (from 11-4-2018 to 11-7-2018, at an average distance $R=0.18$~au). The top panel shows the distribution of the deflection parameter $\mathcal{Z}=1/2(1-\cos\Theta)$ \citep{dudok2020switchbacks}, where $\Theta$ is the angle between the local total magnetic field and the mean field over the integration domain (or over 12 hours for PSP). 
The middle panel compares the scatter plot of $B_x$ vs. the perpendicular components $|{\bf B_\perp}|$ with PSP data (for which we show $-B_r$ on the $y$-axis), after rescaling the numerical solution to 83 nT. The color code shown in the side bar indicates the density of numerical points. The bottom panel shows a 1d cut of $|{\bf B}|$ and $B_x$  at $x=1.13$ and $z=0.0355$ that we compare with a sub-interval of PSP, again after rescaling the numerical solution in amplitude and converting the $y$-axis to the same time interval length of PSP. Fig.~\ref{comparison} shows that the numerical field reproduces essential patterns in the observations: a wide distribution of field deflections including switchbacks ($\mathcal{Z}>0.5$), with the most probable state at a small angle to the radial direction while undergoing sparse deflections at larger angles. However, our solution displays less frequent extreme events and exhibits a smoother profile due to the limited number of modes initialized. Increasing the values of $\alpha$ and $\beta$ and the number of modes could improve the agreement, but stronger tails and smaller scale fluctuations tend to rapidly produce stiff solutions. This suggests that a flatter distribution of ${\mathcal Z}$ might be favored by relaxing the constant-$|{\bf B}|$ constraint.




\section{Discussion}

We have presented a framework to construct 3D magnetic fields that satisfy  the constant-$|\bf {B}|$ constraint exactly, enabling a systematic investigation of longstanding questions about their existence and smoothness. Our method can reproduce broadband fluctuations with embedded switchbacks similar to solar wind observations. However, characteristic analysis and numerical solutions strongly suggest that a smooth field cannot exist in an arbitrarily large spatial domain, as the development of magnetic field discontinuities is generically expected in 3D configurations, when enforcing  the constraint of constant $|{\bf B}|$. Therefore, fluctuations can achieve spherical polarization only within limited regions of space. 
When discontinuities form, the discontinuous continuation of the solution cannot  in  general satisfy both the solenoidal constraint and exactly constant-$|{\bf B}|$. This suggests that a local relaxation of the latter condition is generally required.  

In conclusion, our results indicate that regions of constant $|{\bf B}|$ are organized in limited domains separated by discontinuities associated with magnetic pressure gradients --- a feature that is observed in the solar wind  \citep{ruffolo2021domains}. This interpretation is also consistent with observations of magnetic decreases bounded by rotational discontinuities observed within the Alfvénic spectrum  \citep{tsurutani1994relationship,tsurutani2002relationship,tsurutani2018review,gonzalez2024local}. It may also explain the steeper distribution of ${\mathcal Z}$, when compared with data,  resulting from strictly imposing a constant $|{\bf B}|$. Additionally, our results provide a useful point of comparison with observations of switchback boundaries, which sometimes exhibit characteristics of both rotational and tangential discontinuities~\citep{larosa2021switchbacks}.    
We suggest that the compressive component arising from magnetic pressure gradients is intrinsic to solar wind turbulence when large amplitude, nearly constant-$|{\bf B}|$ fluctuations are present. This points to the importance of incorporating compressible effects into nonlinear Alfvén wave dynamics to explain the dynamical emergence of spherical polarization.

\begin{acknowledgments}
This work was supported by NSF CAREER award~2141564.
\end{acknowledgments}

\bibliography{apssamp}

@article{valentini2019building,
  title={Building up solar-wind-like 3D uniform-intensity magnetic fields},
  author={Valentini, Francesco and Malara, Francesco and Sorriso-Valvo, Luca and Bruno, Roberto and Primavera, Leonardo},
  journal={The Astrophysical Journal Letters},
  volume={881},
  number={1},
  pages={L5},
  year={2019},
  publisher={The American Astronomical Society}
}

@article{roberts2012construction,
  title={Construction of solar-wind-like magnetic fields},
  author={Roberts, D Aaron},
  journal={Physical Review Letters},
  volume={109},
  number={23},
  pages={231102},
  year={2012},
  publisher={APS}
}

@article{squire2022construction,
  title={On the construction of general large-amplitude spherically polarised Alfv{\'e}n waves},
  author={Squire, Jonathan and Mallet, Alfred},
  journal={Journal of Plasma Physics},
  volume={88},
  number={5},
  pages={175880503},
  year={2022},
  publisher={Cambridge University Press}
}

@article{malara1996parametric,
  title={Parametric instability of a large-amplitude nonmonochromatic Alfv{\'e}n wave},
  author={Malara, Francesco and Velli, M},
  journal={Physics of Plasmas},
  volume={3},
  number={12},
  pages={4427--4433},
  year={1996},
  publisher={American Institute of Physics}
}

@article{tenerani2020magnetic,
  title={Magnetic field kinks and folds in the solar wind},
  author={Tenerani, Anna and Velli, Marco and Matteini, Lorenzo and R{\'e}ville, Victor and Shi, Chen and Bale, Stuart D and Kasper, Justin C and Bonnell, John W and Case, Anthony W and de Wit, Thierry Dudok and others},
  journal={The Astrophysical Journal Supplement Series},
  volume={246},
  number={2},
  pages={32},
  year={2020},
  publisher={The American Astronomical Society}
}

@article{marriott2024parametric,
  title={Parametric Instability of Alfv{\'e}n Waves and Wave Packets in Periodic and Open Systems},
  author={Marriott, Maile and Tenerani, Anna},
  journal={The Astrophysical Journal},
  volume={975},
  number={2},
  pages={232},
  year={2024},
  publisher={The American Astronomical Society}
}

@article{primavera2019parametric,
  title={Parametric instability in two-dimensional Alfv{\'e}nic turbulence},
  author={Primavera, Leonardo and Malara, Francesco and Servidio, Sergio and Nigro, Giuseppina and Veltri, Pierluigi},
  journal={The Astrophysical Journal},
  volume={880},
  number={2},
  pages={156},
  year={2019},
  publisher={The American Astronomical Society}
}

@article{shi2024analytic,
  title={Analytic model and magnetohydrodynamic simulations of three-dimensional magnetic switchbacks},
  author={Shi, Chen and Velli, Marco and Toth, Gabor and Zhang, Kun and Tenerani, Anna and Huang, Zesen and Sioulas, Nikos and van der Holst, Bart},
  journal={The Astrophysical Journal Letters},
  volume={964},
  number={2},
  pages={L28},
  year={2024},
  publisher={The American Astronomical Society}
}

@article{barnes1974large,
  title={Large-amplitude hydromagnetic waves},
  author={Barnes, Aaron and Hollweg, Joseph V},
  journal={Journal of Geophysical Research},
  volume={79},
  number={16},
  pages={2302--2318},
  year={1974},
  publisher={Wiley Online Library}
}

@article{matteini2015ion,
  title={Ion kinetic energy conservation and magnetic field strength constancy in multi-fluid solar wind Alfv{\'e}nic turbulence},
  author={Matteini, L and Horbury, TS and Pantellini, F and Velli, M and Schwartz, SJ},
  journal={The Astrophysical Journal},
  volume={802},
  number={1},
  pages={11},
  year={2015},
  publisher={The American Astronomical Society}
}

@article{dudok2020switchbacks,
  title={Switchbacks in the near-Sun magnetic field: long memory and impact on the turbulence cascade},
  author={Dudok de Wit, Thierry and Krasnoselskikh, Vladimir V and Bale, Stuart D and Bonnell, John W and Bowen, Trevor A and Chen, Christopher HK and Froment, Clara and Goetz, Keith and Harvey, Peter R and Jagarlamudi, Vamsee Krishna and others},
  journal={The Astrophysical Journal Supplement Series},
  volume={246},
  number={2},
  pages={39},
  year={2020},
  publisher={The American Astronomical Society}
}

@article{kasper2019alfvenic,
  title={Alfv{\'e}nic velocity spikes and rotational flows in the near-Sun solar wind},
  author={Kasper, Justin C and Bale, Stuart D and Belcher, John W and Berthomier, Matthieu and Case, Anthony W and Chandran, Benjamin DG and Curtis, DW and Gallagher, D and Gary, SP and Golub, L and others},
  journal={Nature},
  volume={576},
  number={7786},
  pages={228--231},
  year={2019},
  publisher={Nature Publishing Group UK London}
}

@article{bale2019highly,
  title={Highly structured slow solar wind emerging from an equatorial coronal hole},
  author={Bale, SD and Badman, ST and Bonnell, JW and Bowen, TA and Burgess, D and Case, AW and Cattell, CA and Chandran, BDG and Chaston, CC and Chen, CHK and others},
  journal={Nature},
  volume={576},
  number={7786},
  pages={237--242},
  year={2019},
  publisher={Nature Publishing Group UK London}
}

@article{huang2025switchbacks,
  title={What are Switchbacks?},
  author={Huang, Zesen and Velli, Marco and Ding, Yuliang},
  journal={arXiv preprint arXiv:2512.12585},
  year={2025}
}

@article{shi2021alfvenic,
  title={Alfv{\'e}nic versus non-Alfv{\'e}nic turbulence in the inner heliosphere as observed by Parker Solar Probe},
  author={Shi, Chen and Velli, Marco and Panasenco, Olga and Tenerani, Anna and R{\'e}ville, Victor and Bale, Stuart D and Kasper, Justin and Korreck, Kelly and Bonnell, JW and de Wit, T Dudok and others},
  journal={Astronomy \& Astrophysics},
  volume={650},
  pages={A21},
  year={2021},
  publisher={EDP Sciences}
}

@article{chen2020evolution,
  title={The evolution and role of solar wind turbulence in the inner heliosphere},
  author={Chen, CHK and Bale, SD and Bonnell, JW and Borovikov, D and Bowen, TA and Burgess, D and Case, AW and Chandran, BDG and de Wit, T Dudok and Goetz, K and others},
  journal={The Astrophysical Journal Supplement Series},
  volume={246},
  number={2},
  pages={53},
  year={2020},
  publisher={The American Astronomical Society}
}

@article{bruno2005observations,
  title={Observations of magnetohydrodynamic turbulence in the 3D heliosphere},
  author={Bruno, R and Carbone, Vincenzo and Bavassano, B and Sorriso-Valvo, La},
  journal={Advances in Space Research},
  volume={35},
  number={5},
  pages={939--950},
  year={2005},
  publisher={Elsevier}
}

@article{belcher1971large,
  title={Large-amplitude Alfv{\'e}n waves in the interplanetary medium, 2},
  author={Belcher, John W and Davis Jr, Leverett},
  journal={Journal of Geophysical Research},
  volume={76},
  number={16},
  pages={3534--3563},
  year={1971},
  publisher={Wiley Online Library}
}

@article{tsurutani1994relationship,
  title={The relationship between interplanetary discontinuities and Alfv{\'e}n waves: Ulysses observations},
  author={Tsurutani, BT and Ho, CM and Smith, EJ and Neugebauer, M and Goldstein, BE and Mok, JS and Arballo, JK and Balogh, A and Southwood, DJ and Feldman, WC},
  journal={Geophysical Research Letters},
  volume={21},
  number={21},
  pages={2267--2270},
  year={1994},
  publisher={Wiley Online Library}
}

@article{tsurutani2018review,
  title={A review of Alfv{\'e}nic turbulence in high-speed solar wind streams: Hints from cometary plasma turbulence},
  author={Tsurutani, Bruce T and Lakhina, Gurbax S and Sen, Abhijit and Hellinger, Petr and Glassmeier, Karl-Heinz and Mannucci, Anthony J},
  journal={Journal of Geophysical Research: Space Physics},
  volume={123},
  number={4},
  pages={2458--2492},
  year={2018},
  publisher={Wiley Online Library}
}

@article{tsurutani2002relationship,
  title={Relationship between discontinuities, magnetic holes, magnetic decreases, and nonlinear Alfv{\'e}n waves: Ulysses observations over the solar poles},
  author={Tsurutani, BT and Galvan, C and Arballo, JK and Winterhalter, D and Sakurai, R and Smith, EJ and Buti, B and Lakhina, GS and Balogh, A},
  journal={Geophysical Research Letters},
  volume={29},
  number={11},
  pages={23--1},
  year={2002},
  publisher={Wiley Online Library}
}

@article{gonzalez2024local,
  title={Local proton heating at magnetic discontinuities in Alfv{\'e}nic and non-Alfv{\'e}nic solar wind},
  author={Gonz{\'a}lez, CA and Verniero, JL and Bandyopadhyay, R and Tenerani, A},
  journal={The Astrophysical Journal},
  volume={963},
  number={2},
  pages={148},
  year={2024},
  publisher={The American Astronomical Society}
}

@article{larosa2021switchbacks,
  title={Switchbacks: statistical properties and deviations from Alfv{\'e}nicity},
  author={Larosa, A and Krasnoselskikh, V and de Wit, T Dudok and Agapitov, O and Froment, C and Jagarlamudi, VK and Velli, M and Bale, SD and Case, AW and Goetz, K and others},
  journal={Astronomy \& Astrophysics},
  volume={650},
  pages={A3},
  year={2021},
  publisher={EDP Sciences}
}

@article{goldstein1974, title={On the theory of large amplitude Alfven waves}, author={Goldstein, M. and Klimas,  A. J. and Barish, F. D.}, journal={Proc. of the Solar Wind 3 Conf.}, year={1974}
}

@article{barnes1976nonexistence,
  title={On the nonexistence of plane-polarized large amplitude Alfv{\'e}n waves},
  author={Barnes, Aaron},
  journal={Journal of Geophysical Research},
  volume={81},
  year={1976}
}

@article{ruffolo2021domains,
  title={Domains of magnetic pressure balance in Parker Solar Probe observations of the solar wind},
  author={Ruffolo, David and Ngampoopun, Nawin and Bhora, Yash R and Thepthong, Panisara and Pongkitiwanichakul, Peera and Matthaeus, William H and Chhiber, Rohit},
  journal={The Astrophysical Journal},
  volume={923},
  number={2},
  pages={158},
  year={2021},
  publisher={The American Astronomical Society}
}

@article{sioulas23,
title={On the evolution of the anisotropic scaling of magnetohydrodynamic turbulence in the inner heliosphere},
author={Sioulas, Nikos and Marco, Velli and Zesen, Huang and Chen, Shi and Trevor A., Bowen and B. D. G., Chandran et al.},
journal={The Astrophysical Journal},
volume={951},
number={2},
pages={141},
year={2023}
}

@article{horbury2012anisotropy,
  title={Anisotropy in space plasma turbulence: solar wind observations},
  author={Horbury, TS and Wicks, RT and Chen, CHK},
  journal={Space Science Reviews},
  volume={172},
  number={1},
  pages={325--342},
  year={2012},
  publisher={Springer}
}

@article{dasso2005anisotropy,
  title={Anisotropy in fast and slow solar wind fluctuations},
  author={Dasso, S and Milano, LJ and Matthaeus, WH and Smith, CW},
  journal={The Astrophysical Journal Letters},
  volume={635},
  number={2},
  pages={L181--L184},
  year={2005}
}

\end{document}